\documentclass{cupconf}
\usepackage{natbib} \bibpunct{(}{)}{;}{a}{}{,}
\usepackage{graphicx}

\title{Are WNL stars tracers of high metallicity?}

\author[G.\ Gr\"afener \& W.-R.\ Hamann]{G.\ Gr\"afener$^1$ \& W.-R.\ Hamann$^1$}

\affiliation{$^1$Institut f\"ur Physik, Universit\"at Potsdam, Am Neuen Palais
  10, 14469 Potsdam, Germany}

\pubyear{2006}
\volume{XXX}
\pagerange{XXX}
\date{}
\begin{document}

\maketitle

\begin{abstract}
  We present new atmosphere models for Wolf-Rayet stars that include a
  self-consistent solution of the wind hydrodynamics. We demonstrate that the
  formation of optically thick WR winds can be explained by radiative driving
  on Fe line opacities, implying a strong dependence on metallicity ($Z$).
  $Z$-dependent model calculations for late-type WN stars show that these
  objects are very massive stars close to the Eddington limit, and that their
  formation is strongly favored for high metallicity environments.
\end{abstract}

\firstsection 
\section{PoWR hydrodynamic model atmospheres}

The Potsdam Wolf-Rayet (PoWR) hydrodynamic model atmospheres combine fully
line-blanketed non-LTE models with the equations of hydrodynamics \citep[for
details see][]{gra1:05,ham1:03,koe1:02,gra1:02}. The wind structure ($\rho(r)$
and $v(r)$) and the temperature structure $T(r)$ are computed in line with the
full set of non-LTE populations, and the radiation field in the co-moving
frame (CMF). In contrast to all previous approaches, the radiative wind
acceleration $a_{\rm rad}$ is obtained by direct integration
\begin{equation}
 \label{eq:arad}
  a_{\rm rad} = \frac{1}{c} \int \chi_\nu F_\nu {\rm d}\nu,
\end{equation}
instead of making use of the Sobolev approximation.  In this way, complex
processes like strong line overlap, or the redistribution of radiation, are
automatically taken into account. Moreover, the models include small-scale
wind clumping \citep[throughout this work we assume a clumping factor of
$D=10$, for details see also][]{ham1:98}. The models describe the conditions
in WR\,atmospheres in a realistic manner, and provide synthetic spectra, i.e.\ 
they allow for a direct comparison with observations.

Utilizing these models, we recently obtained the first fully self-consistent
Wolf-Rayet wind model, for the case of an early-type WC star with strong lines
\citep{gra1:05}.  Moreover, we have examined the mass loss from late-type WN
stars and its dependence on metallicity \citep{gra3:06,gra1:06}.

\section{Spectral analyses of galactic WR stars}

\begin{figure}
\center{\includegraphics[scale=0.55]{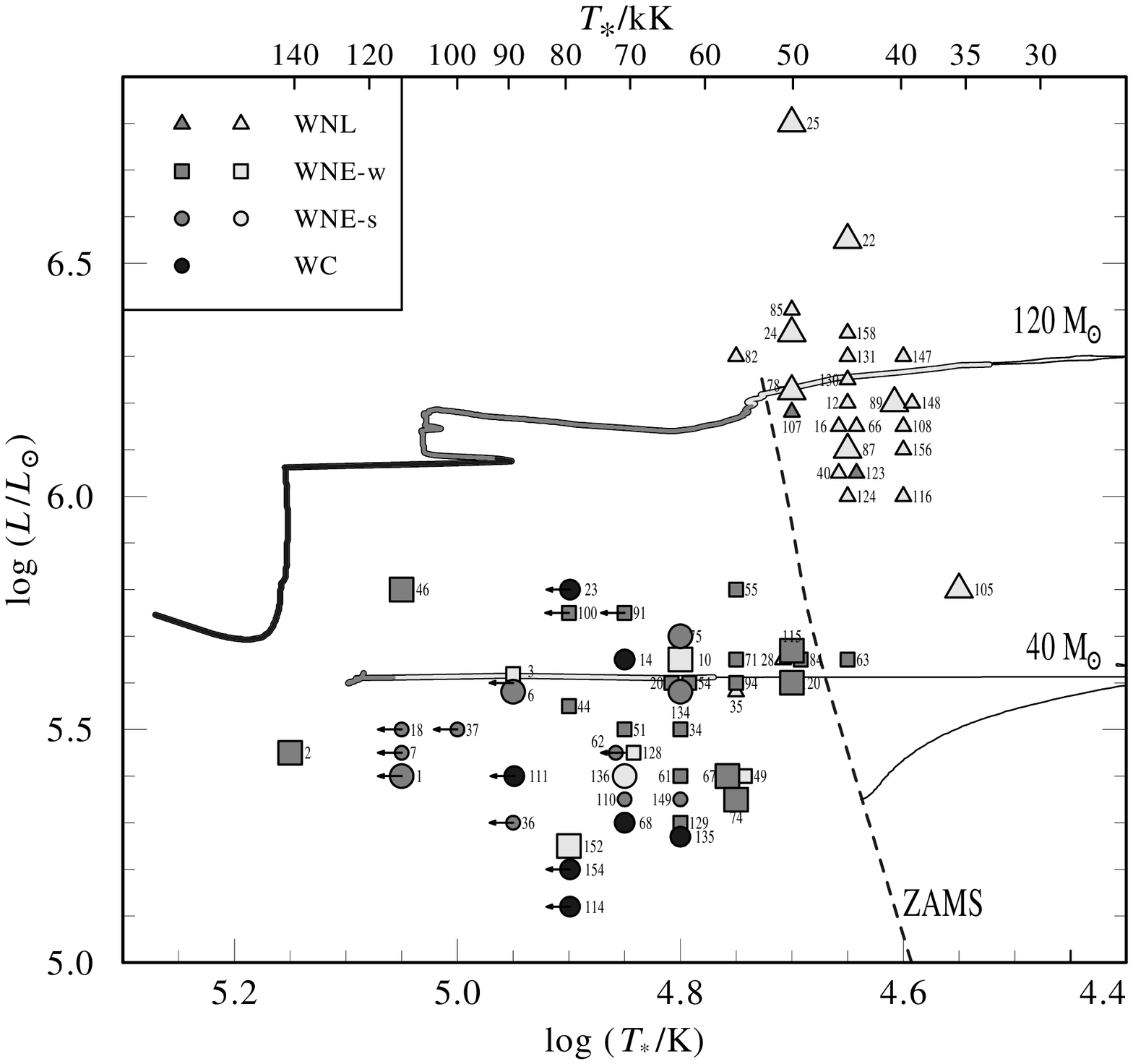}}
\caption{Recent spectral analyses of galactic WR stars with line-blanketed
  models, according to \citet{ham1:06} and \citet{bar2:06}: symbols in light
  grey denote H-rich WR stars, whereas H-free objects are indicated in dark
  grey.  WC stars are indicated in black.  For objects with large symbols
  distance estimates are available \citep{huc1:01}, whereas objects
  with small symbols are calibrated by their spectral subtype.  Evolutionary
  tracks for non-rotating massive stars \citep{mey1:03} are shown for
  comparison.
  \label{fig:hrd}
}
\end{figure}

A comprehensive study of galactic WR stars, based on spectral analyses with
line-blanketed PoWR models (\citet{ham1:06} for WN\,stars, \citet{bar2:06} for
WC\,stars), revealed a bimodal WR subtype distribution in the HRD, where the
H-rich WNL\,stars are located to the right of the ZAMS with luminosities above
$10^6\,L_\odot$, whereas the (mostly H-free) early- to intermediate WN
subtypes, as well as the WC stars, show lower luminosities and hotter
temperatures (see Fig.\,\ref{fig:hrd}).

This dichotomy already implies that the H-rich WNL\,stars are the descendants
of very massive stars, possibly still in the pase of central H-burning,
whereas the earlier subtypes (including the WC stars) are more evolved, less
massive, He-burning objects. Note, however, that distance estimates are
only available for a small part of the WNL sample. Some of these objects thus
might have lower luminosities and be the direct progenitors of the earlier
subtypes.

\section{Hydrodynamic atmosphere models for WNL stars}

\begin{figure}
\center{\includegraphics[scale=0.45]{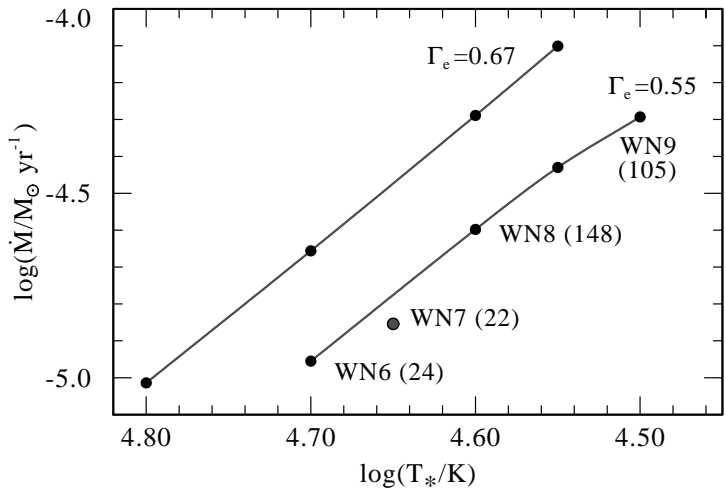}}
\caption{Wind models for galactic WNL stars:
  mass loss rates for different stellar temperatures $T_\star$ and Eddington
  factors $\Gamma_{\rm e}$. The corresponding spectral subtypes are indicated
  together with WR numbers of specific galactic objects \citep[according to][
  in brackets]{huc1:01}, showing a good agreement with the synthetic
  line spectra.  Note that the models are computed for a fixed stellar
  luminosity of $10^{6.3}\,L_\odot$.  The WN\,7 model (WR\,22) is slightly
  offset from the standard grid models because it is calculated with an
  enhanced hydrogen abundance (see text).
  \label{mdot-wnl}
}
\end{figure}

In a recent work we have investigated the properties of the luminous, H-rich
WNL stars with our hydrodynamic PoWR models \citep{gra3:06}. The most
important conclusion from that work is that WR-type mass loss is primarily
triggered by high $L/M$ ratios or, equivalently, Eddington factors
$\Gamma_{\rm e}\equiv\chi_{\rm e}L_\star/4\pi c G M_\star$ approaching unity.
Note that high $L/M$ ratios are expected for very massive stars {\em and} for
He-burning objects, giving a natural explanation for the occurrence of the
WR phenomenon.

In Fig.\,\ref{mdot-wnl} we show the results from grid computations for WNL
stars with a fixed luminosity of $10^{6.3}\,L_\odot$, and stellar temperatures
$T_\star$ in the range of 30--60\,kK.  For the stellar masses, values of 67
and $55\,M_\odot$ are adopted, corresponding to Eddington factors of
$\Gamma_{\rm e}$\,$=$\,0.55 and 0.67.
Notably, the mass loss strongly depends on $\Gamma_{\rm e}$ and $T_\star$.
The obtained synthetic spectra nicely reflect the observed sequence of {\em
  weak-lined} WNL subtypes, starting with WN\,6 at 55\,kK to WN\,9 at 31\,kK.
From a more detailed investigation of the WN\,7 component in WR\,22, an
eclipsing WR+O binary system in Car\,OB1, we infer a stellar mass of
$78\,M_\odot$ ($\Gamma_e=0.67$), in agreement \citet{rau1:96} who obtained
$72\pm 3\,M_\odot$ from the binary orbit.
Such high stellar masses imply that the weak-lined WNL stars are still in the
pase of central H-burning, suggesting an evolutionary sequence of the form \,O
$\rightarrow$ WNL $\rightarrow$ LBV $\rightarrow$ WN $\rightarrow$ WC\, for
very massive stars.

\section{WNL stars at different metallicities}

\begin{figure}
  \center{\includegraphics[scale=0.4]{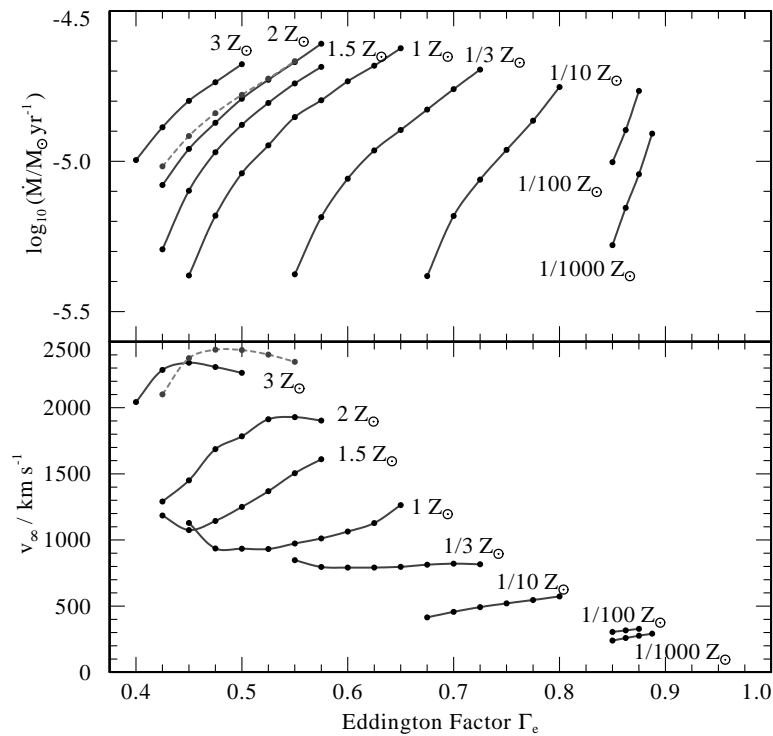}}
\caption{WNL star mass loss over a broad range of metallicities ($Z$): Mass
  loss rates (top) and terminal wind velocities (bottom), as obtained from our
  hydrodynamic models, are plotted vs.\ the Eddington factor $\Gamma_{\rm e}$.
  The solid curves indicate model series for WNL stars where $\Gamma_{\rm e}$
  is varied for a given value of $Z$.  The dashed grey lines indicate models
  with $3\,Z_\odot$ and $X_{\rm H}=0.7$, corresponding to very massive, metal
  rich stars on the ZAMS.
\label{wnl-z}
}
\end{figure}

In Fig.\,\ref{wnl-z} we present results from a grid of models for luminous WNL
stars at different metallicities \citep{gra3:06}.  The models are computed for
a fixed luminosity of $10^{6.3}\,L_\odot$, a stellar temperature of
$T_\star=45\,$kK, and a hydrogen surface mass fraction of $X_{\rm H}=0.4$.  In
addition to the metal abundances we have varied the Eddington factor
$\Gamma_e$ (or equivalently the stellar mass). Note that we have scaled all
metals with $Z$, assuming a CNO-processed (i.e.\ N-enriched) WN surface
composition. The wind driving in our models is chiefly due to radiation
pressure on Fe-group line opacities. In accordance with \citet{vin1:05} we
thus find a strong dependence on $Z$. However, as in our previous
computations, the proximity to the Eddington limit plays an equally important
role. We find that optically thick winds with high WR-type mass loss rates are
formed over the whole range of metallicities, from 1/1000--3\,$Z_\odot$, if
the stars get close enough to the Eddington limit. Only the limiting value of
$\Gamma_e$ where the WR-type winds start to form, changes. For solar $Z$,
values of $\Gamma_e = 0.5$--0.6 are leading to the formation of weak-lined WNL
stars.  As we have seen for the case of WR\,22, this corresponds to very
massive, slightly over-luminous stars in a late phase of H-burning. Note that
it is indeed observed that the most massive stars in very young galactic
clusters are in the WNL phase (e.g., \citet{fig1:02}, \citet{naj1:04} for the
Arches cluster; \citet{dri1:99}, \citet{cro1:98} for NGC\,3603).

For higher values of $Z$, the limit for the formation of WR-type winds shifts
towards even lower values of $\Gamma_e$. For $3\,Z_\odot$, we find that stars
with $\Gamma_e \approx 0.4$ already show typical WNL mass loss rates (see
Fig.\,\ref{wnl-z}). This corresponds to objects with $120\,M_\odot$ on the
ZAMS. We thus expect that metal rich stars with very high masses already start
their life in the WNL phase, i.e., the occurrence of WNL stars in young
massive clusters is strongly favored for high metallicities.

\newpage


\begin{thebibliography}{17}
\expandafter\ifx\csname natexlab\endcsname\relax\def\natexlab#1{#1}\fi

\bibitem[{{Barniske} {et~al.}(2006){Barniske}, {Hamann}, \&
  {Gr{\"a}fener}}]{bar2:06}
{Barniske}, A., {Hamann}, W.-R., \& {Gr{\"a}fener}, G. 2006, in {Stellar
  Evolution at Low Metallicity: Mass-Loss, Explosions, Cosmology}, ed.
  H.~{Lamers}, N.~{Langer}, \& T.~{Nugis}, ASP Conference Series, in press

\bibitem[{Crowther \& Dessart(1998)}]{cro1:98}
Crowther, P.~A. \& Dessart, L. 1998, MNRAS, 296, 622

\bibitem[{Drissen(1999)}]{dri1:99}
Drissen, L. 1999, in IAU Symp., Vol. 193, Wolf-{R}ayet phenomena in massive
  stars and starburst galaxies, ed. K.~A. van~der Hucht, G.~Koenigsberger, \&
  P.~R.~J. Eenens, 403

\bibitem[{Figer {et~al.}(2002)Figer, Najarro, Gilmore, Morris, Kim, Serabyn,
  McLean, Gilbert, Graham, Larkin, Levenson, \& Teplitz}]{fig1:02}
Figer, D.~F., Najarro, F., Gilmore, D., {et~al.} 2002, ApJ, 581, 258

\bibitem[{{Gr{\" a}fener} \& {Hamann}(2005)}]{gra1:05}
{Gr{\" a}fener}, G. \& {Hamann}, W.-R. 2005, A\&A, 432, 633

\bibitem[{{Gr{\" a}fener} \& {Hamann}(2006a)}]{gra3:06}
{Gr{\" a}fener}, G. \& {Hamann}, W.-R. 2006a, A\&A, submitted

\bibitem[{{Gr{\"a}fener} \& {Hamann}(2006b)}]{gra1:06}
{Gr{\"a}fener}, G. \& {Hamann}, W.-R. 2006b, in {Stellar Evolution at Low
  Metallicity: Mass-Loss, Explosions, Cosmology}, ed. H.~{Lamers}, N.~{Langer},
  \& T.~{Nugis}, ASP Conference Series, in press

\bibitem[{{Gr\"afener} {et~al.}(2002){Gr\"afener}, {Koesterke}, \&
  {Hamann}}]{gra1:02}
{Gr\"afener}, G., {Koesterke}, L., \& {Hamann}, W.-R. 2002, A\&A, 387, 244

\bibitem[{{Hamann} \& {Gr{\"a}fener}(2003)}]{ham1:03}
{Hamann}, W.-R. \& {Gr{\"a}fener}, G. 2003, A\&A, 410, 993

\bibitem[{{Hamann} {et~al.}(2006){Hamann}, {Gr{\"a}fener}, \&
  {Liermann}}]{ham1:06}
{Hamann}, W.-R., {Gr{\"a}fener}, G., \& {Liermann}, A. 2006, A\&A, in press

\bibitem[{{Hamann} \& {Koesterke}(1998)}]{ham1:98}
{Hamann}, W.-R. \& {Koesterke}, L. 1998, A\&A, 335, 1003

\bibitem[{{Koesterke} {et~al.}(2002){Koesterke}, {Hamann}, \& {Gr{\"
  a}fener}}]{koe1:02}
{Koesterke}, L., {Hamann}, W.-R., \& {Gr{\" a}fener}, G. 2002, A\&A, 384, 562

\bibitem[{{Meynet} \& {Maeder}(2003)}]{mey1:03}
{Meynet}, G. \& {Maeder}, A. 2003, A\&A, 404, 975

\bibitem[{{Najarro} {et~al.}(2004){Najarro}, {Figer}, {Hillier}, \&
  {Kudritzki}}]{naj1:04}
{Najarro}, F., {Figer}, D.~F., {Hillier}, D.~J., \& {Kudritzki}, R.~P. 2004,
  ApJ, 611, L105

\bibitem[{{Rauw} {et~al.}(1996){Rauw}, {Vreux}, {Gosset}, {Hutsemekers},
  {Magain}, \& {Rochowicz}}]{rau1:96}
{Rauw}, G., {Vreux}, J.-M., {Gosset}, E., {et~al.} 1996, A\&A, 306, 771

\bibitem[{{van der Hucht}(2001)}]{huc1:01}
{van der Hucht}, K.~A. 2001, New Astronomy Review, 45, 135

\bibitem[{{Vink} \& {de Koter}(2005)}]{vin1:05}
{Vink}, J.~S. \& {de Koter}, A. 2005, A\&A, 442, 587

\end{thebibliography}

\end{document}